\documentclass[epj]{svjour}
\usepackage{graphics}
\begin{document}
\title{Exclusive and semi-inclusive strangeness and
charm production in $\pi N$ and $NN$ reactions \thanks{
Supported by DFG, RFFI and Forschugszentrum J\"ulich.}}
\author{A.M.~Gasparyan\inst{1,2}, V.Yu.~Grishina\inst{3},
L.A.~Kondratyuk\inst{1,2}, W.~Cassing\inst{4}, J.~Speth\inst{1}}
\institute{IKP, Forschungszentrum J\"ulich, D-52425 J\"ulich, Germany \and
Institute for Theoretical and Experimental Physics, B.~Cheremushkinskaya 25,
 117259 Moscow, Russia\and Institute for Nuclear Research, 60th October
Anniversary Prospect 7A, 117312 Moscow, Russia\and  Institute for Theoretical Physics,  University of
Giessen, Heinrich-Buff-Ring 16, D-35392 Giessen, Germany }
\date{Received: date / Revised version: date}

\abstract{Using the Quark-Gluon Strings Model (QGSM) combined with
Regge phenomenology we consider the reactions $\pi^- p \to K^0
\Lambda$ and $\pi^- p \to D^- \Lambda_c^+$ which are dominated by
the contributions of the  $K^*$ and $D^*$ Regge trajectories,
respectively. The spin structure of the amplitudes is  described
by introducing Reggeized Born terms.  It is found that the
existing data for the reaction  $\pi^- p \to K^0 \Lambda$ are in
reasonable agreement with the model predictions.  To describe the
absolute values of the cross sections it is necessary to introduce
also suppression factors which can be related to absorption
corrections. Furthermore, assuming the SU(4) symmetry to hold for
Regge residues and the universality of absorption corrections we
calculate the cross section of the reaction $\pi^- p \to  D^-
\Lambda_c^+$.  Employing the latter results from $\pi^- p$
reactions we then estimate the contributions of the pion exchange
mechanism to the cross sections of the reactions $NN \to
NK\Lambda$ and $NN \to N\bar D \Lambda_c$ and compare them with
the contributions of the $K$ and $D$ exchanges.  We find that the
$NN$ reactions are dominated not by pion exchange but by $K$ and
$D$ exchanges, respectively. Moreover, assuming the SU(4) symmetry
to hold approximately for the coupling constants $g_{ND
\Lambda_c}$ = $g_{NK \Lambda}$ we analyze also the production of
leading $\Lambda_c$ hyperons in the reaction $NN \to \Lambda_c X$.
It is shown that the non-perturbative mechanism should give an
essential contribution to the $\Lambda_c$ yield for $x \geq 0.5$.}

%
\authorrunning{A. M. Gasparyan et al.}

\titlerunning{Exclusive and semi-inclusive strangeness and charm production...}

\maketitle

Recently it has been argued  \cite{Cassing2001} that the open
charm enhancement observed in nucleus-nucleus collisions
\cite{Abreu2000} at the SPS might be due to secondary reaction
mechanisms such as $\pi N\to \bar D \Lambda_c$ or $NN \to N\bar D
\Lambda_c$.  In this work we present estimates of these elementary
cross sections  using the anology with strangeness production in
$\pi N$ and $NN$ collisions.  We consider also semi-inclusive $
\Lambda$ and $\Lambda_c$ production in the reactions $NN \to N X
\Lambda$ and $NN \to N X \Lambda_c$.

It is well known that the methods of perturbative QCD can not be
applied for a calculation of the cross sections mentioned above
especially at invariant energies closer to threshold. For the
analysis of binary reactions we instead use the nonperturbative
Quark-Gluon String model \cite{Kaidalov82} and for reactions with
three particles in the final state we employ the meson-exchange
model taking into account the exchanges of the lowest meson states
- pseudoscalar and vector.

\begin{figure}
\resizebox{0.45\textwidth}{!} {%
\includegraphics{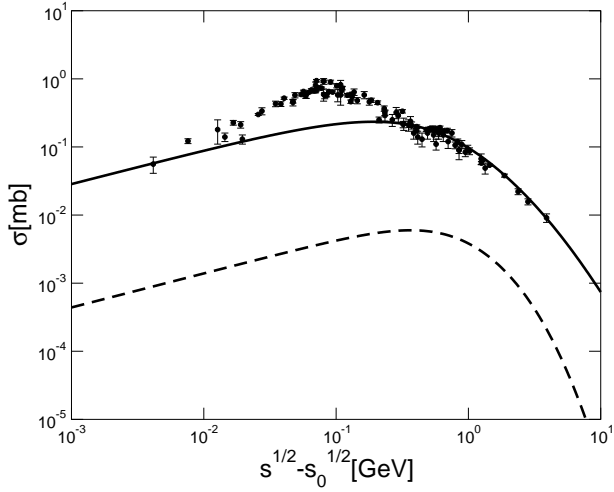}
} \caption{Total cross section for the reaction $\pi^- p \to K^+
\Lambda$ (solid line) and $\pi^- p\to D^-\Lambda_c^+$ (dashed
line) as a function of the invariant energy above thresholds in
comparison to the data from Ref.  \protect\cite{Baldini}.}
\label{piN}
\end{figure}

The amplitudes for the reactions $\pi N\to K\Lambda$ and $\pi N\to
D\Lambda_c$ are calculated using the Reggeized Born term approach
(see e.g. Refs. \cite{Irving,Guidal}) with contributions of $K^*$
and $D^*$ Regge trajectories, respectively.  The parameters of the
trajectories are taken from  Ref. \cite{Boreskov}, whereas for the
coupling constants we assume SU(4) symmetry as suggested recently
by Lin and Ko \cite{Lin}.  With these parameters the energy
dependence of the total $\pi^-p\to K^+\Lambda$ cross section
(solid line in Fig. \ref{piN}) as well as the $t$-dependence of
the differential $\pi^- p\to K^+\Lambda$ cross section are
described rather well, except for the region close to threshold
where the dominant contribution stems from the well established
$s$- and $p$-wave resonances \cite{Feuster}. We note that to
obtain the absolute value of the cross section one has to
introduce a suppression factor of  $\sim 0.4$, which can be
interpreted as an absorption correction. Assuming its universality
we will introduce the same suppression factor for  charm
production, too. The resulting total cross section of the reaction
$\pi^- p\to D^-\Lambda_c^+$ is shown by the dashed line in Fig.
\ref{piN}.

Our next step is to study $NN$ reactions where we first apply our
model to strangeness production.  Using the method of Yao
\cite{Yao} one can express the $\pi$-exchange cross section for
the reaction $pp \to pK^+\Lambda$ in terms of the $g_{NN \pi}$
coupling constant and the $\pi^0 p \to K^+ \Lambda$ cross section
as
\begin{eqnarray}
&&\nonumber\sigma =\frac {g^2_{NN \pi}} {8 {\pi}^2 p_i^2 s} \int
_{W_{min}}^{W_{max}} k \  W^2 \ \sigma (\pi^0 p \to K^+ \Lambda,
W) \ dW \times \nonumber \\ &&\times \int
_{t_{min}(W)}^{t^{max}(W)} F^4_\pi(t) \frac{1}{(t-m_\pi^2)^2} \ t
\ dt,\label{fyao}
\end{eqnarray}
where $W$ is the c.m. energy in the  $K^+ \Lambda$ subsystem and
$t$ is the 4-momentum transfer squared  between the initial and
final baryons. The form factor was chosen to be of the standard
monopole type:
$F_\pi(t)=(\Lambda_\pi^2-m_\pi^2)/(\Lambda_\pi^2-t)$ with
$\Lambda_\pi=1.3$ GeV.  A similar expression -- but with
$\sigma(K^+p)$ -- can be written for the $K$-exchange. For the
$K^+p$ elastic cross section employed in this case we use the
parametrization of Cugnon et al.  \cite{Cugnon}.

\begin{figure}
\resizebox{0.45\textwidth}{!}{%
  \includegraphics{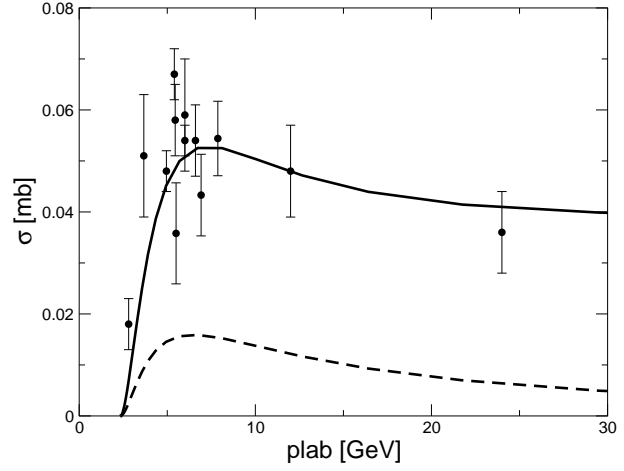}
} \caption{The total cross section for the reaction $p p \to K^+
\Lambda p$ as a function of proton laboratory momentum $p_{lab}$.
The dashed line denotes the $\pi$-exchange contribution while the
solid line corresponds to the $K$-exchange with the cutoff
$\Lambda_{K}$=1.0 GeV} \label{pppkl}
\end{figure}

The total cross section of the reaction $p p \to K^+ \Lambda p$ as
a function of the laboratory momentum $p_{lab}$ is shown in Fig.
\ref{pppkl}. The dashed and solid lines describe the $\pi$- and
$K$-exchange contributions, respectively, with the cutoff
$\Lambda_{K}$ = 1.0 GeV.   An interesting observation is that the
pion-exchange contribution is substantially smaller than the
$K$-exchange and can be neglected especially at higher energies.
The reason for that is a difference in the energy dependence of
the elementary cross sections: $\sigma(\pi^-p\to K\Lambda)$ falls
off with energy whereas $\sigma(K^+p)$ is almost constant since it
is dominated by Pomeron exchange. Moreover, the $K$-exchange alone
is able to reproduce the experimental data when choosing the
cutoff $\Lambda_K=1.0$ GeV.

We see that using the approach of Yao \cite{Yao} we can express
the cross section for the reaction $pp \to K^+ p \Lambda $ through
the coupling constant $g_{ K^+ p \Lambda }$ and the elastic $K^+
p$ scattering cross section $\sigma_{el} (K^+p)$. Similarly, the
cross section for the leading $\Lambda$ production in the reaction
$pp \to X \Lambda $ can be expressed through the same coupling
constant and the total $K^+ p$ scattering cross section
$\sigma_{tot} (K^+p)$.

As follows from Fig.~2 the cross section of the reaction  $pp \to
K^+ p \Lambda$ is about 40--50 $\mu$b for $p_{lab} \geq 5$GeV/c.
The ratio of the cross sections $\sigma_{tot} (K^+p)/ \sigma_{el}
(K^+p) \simeq 7-8$.  Thus we expect that the cross section for
semi-inclusive leading $\Lambda$ production in the reaction $pp
\to X \Lambda $ via $K$ exchange should be about
$\sigma_{K-exch}(pp \to X \Lambda) \simeq 300-400~ \mu$b.
Furthermore, the ratio of the coupling constants $g_{ K^{*+} p
\Lambda }/ g_{ K^+ p \Lambda } \simeq 2$ \cite{Guidal} which
implies that the contribution of the $K^*$ exchange to the cross
section of the leading $\Lambda$ production might be $\sim $4
times larger. Thus we expect the cross section for the
semi-inclusive leading $\Lambda$ production to be about
$$\sigma_{K-exch}(pp \to X \Lambda) + \sigma_{K*-exch}(pp \to X
\Lambda) \simeq 1.5 \div 2 \rm{mb}.$$ We note that Erhan et al.
\cite{Erhan} quote total cross sections for the reaction $pp \to
\Lambda+X$ of $4.4 \pm 0.2$  and $4.7 \pm 0.2$ mb at $\sqrt{s}=53$
and 62 GeV, respectively. This comparison shows that the
 mechanism considered above  gives a  dominant contribution to the
semi-inclusive leading $\Lambda$ production in the reaction $pp
\to X \Lambda $.

Using the analogy  of strangeness and charm production we can
expect that the main contributions to the cross sections of the
reactions $NN \to \bar{D}_c (\bar{D}_c^*) \Lambda_c N$ come from
the $D_c$ and $D_c^*$ exchanges, respectively.  The coupling
constants -- involving a charm quark -- can be related to the
strange ones using $SU(4)$ symmetry, i.e. $g_{ K N \Lambda }=g_{
D_c N \Lambda _c} $ and $g_{ K^{*+} N \Lambda }=g_{ D_c^{*} N
\Lambda_c } $. Within the approach of Yao \cite{Yao} we then can
express the cross section of the reaction $pp \to \bar{D}_c^0 p
\Lambda_c^{+}$ through the coupling constant $g_{ \bar{D}_c^0 p
\Lambda_c^+ }$ and the elastic $\bar{D}_c^0 p$ scattering cross
section $\sigma_{el} (\bar{D}_c^0 p)$. Similarly, the cross
section for the leading $\Lambda_c$ production in the reaction $pp
\to X \Lambda_c $  can be expressed through the same coupling
constant and the total $\bar{D}_c^0 p$ scattering cross section
$\sigma_{tot} (\bar{D}_c^0 p)$.

\begin{figure}
\resizebox{0.45\textwidth}{!}{%
  \includegraphics{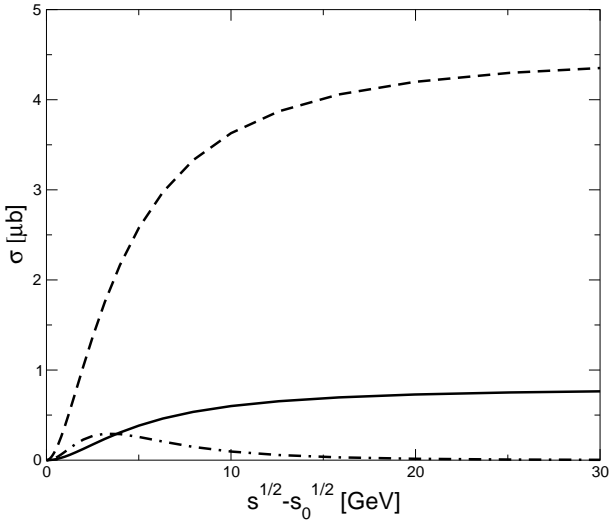}
} \caption{The predicted  total cross section for the reaction $p
p \to \bar D^0 \Lambda_c^+ p$ as a function of the invariant
energy above threshold. The dash-dotted line denotes the
contribution from the $\pi$-exchange while the solid line and the
dashed line correspond to the $D_c$-exchange with the cutoff
$\Lambda_D=1.0$ GeV and $1.5$ GeV, respectively.}
\label{pppdl}
\end{figure}

In our calculations we assume $\sigma_{el} (\bar{D}_c^0 p)$ =
$\sigma_{el} (K^+p)$ and $\sigma_{tot}
 (\bar{D}_c^0 p)$ = $\sigma_{tot} (K^+p)$ while the form factor is taken as
 \begin{equation}
 \label{ff1} F_D(t) = \Lambda_D^2/( \Lambda_D^2-t). \end{equation}
 In Fig. \ref{pppdl} we
present the total cross section for the reaction $p p \to
\bar{D}_c^0 \Lambda_c^+ p$  as a function of the invariant energy
above threshold for $\Lambda_{D_c}$=1.5 GeV (dashed line) and
$\Lambda_{D_c}$=1.0 GeV (solid line). The dash-dotted line denotes
the contribution from the $\pi$-exchange alone.  Note, that for
the elementary reaction $\pi^0p\to \bar D^0 \Lambda^+_c$ we use
the amplitude calculated in the approach discussed above, while
for the $\bar D^0 p$ cross section we adopt a value corresponding
to the asymptotic $K^+ p$ cross section, i.e. $\sim 3$ mb, which
is consistent with the values used in the literature (see e.g.
\cite{Cassing00}).

We find that the main contribution to the cross section for the
reaction $NN \to \bar D \Lambda_c N$ (a few GeV above threshold)
comes from the $D_c$ exchange which is much larger than the pion
exchange for  cut-off parameters $\Lambda_D \geq $ 1 GeV.

\begin{figure}
\resizebox{0.45\textwidth}{!}{%
  \includegraphics{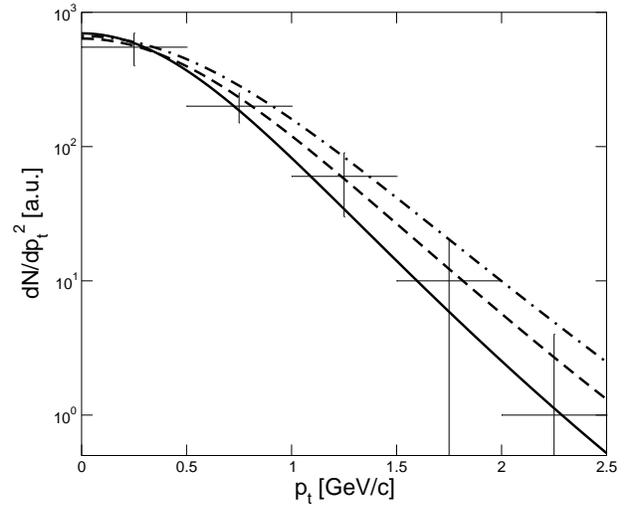}
}
\caption{The $p_T$ dependence of the differential cross section
for the reaction $pp \to X \Lambda_c $ at $\sqrt{s}=62$ GeV. The
theoretical curves correspond to the differential cross section
$d^2\sigma/dp_t^2dx$ calculated at $x=0.9$ (where it peaks) for
the cutoffs $\Lambda_D=1.0$ GeV (solid line), $\Lambda_D=1.3$ GeV
(dashed line), $\Lambda_D=1.5$ GeV (dash-dotted line). The results
are normalized to the data from \protect\cite{Bari} at small
transverse momentum $p_t$. } \label{dsig}
\end{figure}

To find restrictions on the cutoff parameter $\Lambda_D$ in
(\ref{ff1}) we use the data from Ref. \cite{Bari} on
semi-inclusive $ \Lambda_c$ production in the reaction $pp \to X
\Lambda_c$.  We assume now that the same $D$-exchange mechanism
also gives a large contribution to the semi-inclusive $ \Lambda_c$
production at $x$ close to 1. (In fact, in our calculation the
cross section is peaked at $x\sim 0.9$).  Of course, in this case
one has to insert the total $\bar D^0 p$ cross section in the
corresponding analog of Eq. (\ref{fyao}). As shown in Fig.
\ref{dsig} the $p_t$ dependence of the differential cross section
constrains $\Lambda_D$ to $1-1.5$ GeV.

To make a rough estimate of the absolute value of the $D$-exchange
contribution to the leading $\Lambda_c$ production in the reaction
$pp \to X \Lambda_c $ we assume that  the total $\bar D^0 p$ cross
section is the same as in case of $K^+ p$ scattering, i.e. $\sim
20$ mb. Then at c.m. energies larger than 10 GeV we obtain a cross
section of $\sim10-40~ \mu b$  depending on the choice of the
cutoff. As in the case of strangeness production the contribution
from $D^*$ exchange might be approximately 4 times larger.
Therefore, according to our estimates  the cross section of the
semi-inclusive leading  $\Lambda_c$ production at high energy
should be as large as $\sim 50-200~ \mu b$. This estimate agrees
with the experimental value of $40-200~ \mu b$ at $\sqrt{s}$=62
GeV quoted in Ref. \cite{Bari} which implies that the
non-perturbative mechanism considered here gives an essential
contribution to the leading $\Lambda_c$ production.

We finally note, that the same mechanism with $D$ and $D^*$
exchanges should provide a similar contribution to the open charm
production in $p \bar p$ collisions.

\end{document}